# Efficient Equalization Method for Cyclic Prefix-Free Coarsely Quantized Massive MIMO Systems

Jawad Munir, Daniel Plabst, and Josef A. Nossek
Methods of Signal Processing, Technische Universität München, 80290 Munich, Germany
Email: {jawad.munir, daniel.plabst, josef.a.nossek}@tum.de

***Abstract*—The use of low resolution *Analog to Digital Converters* (ADCs) can significantly reduce the power consumption for massive *Multiple Input Multiple Output* (MIMO) systems. The existing literature on quantized massive MIMO systems deals with *Cyclic Prefix* (CP) transmission schemes in frequency-selective fading channels. In this paper, we propose a block processing *Frequency Domain Equalization* (FDE) technique in CP-free transmission schemes for massive MIMO systems having low resolution ADCs. The optimal block length for FDE is found by minimizing a computational complexity cost function and taking quantization distortion, channel impulse response and the number of transmit and receiver antennas into account. Through numerical simulation, it is shown that the optimal block length also guarantees good performance in terms of the *Mean Square Error* (MSE) and *Bit Error-Rate* (BER) criterion.**

## I. Introduction

The next generation of wireless communication systems promises to increase the spectral efficiency by deploying a very large number of antennas at the *Base Station* (BS), a concept known as massive MIMO [1]. A major concern for the adoption of massive MIMO systems is the prohibitive increase of power consumption in the *Radio Frequency* (RF)-chain due to the large number of used antennas. Survey [2] shows that the ADC is a bottleneck, as the power consumption scales roughly exponentially in the number of quantization bits. Therefore, low resolution ADCs, even down to a single bit, have gained a lot of interest in massive MIMO literature, as they provide effective means to reduce the overall power consumption of the RF-chain.

Most previous work on massive MIMO systems with low resolution ADCs deals with frequency-flat channels [3], [4]. The algorithms for channel estimation and data detection have recently been proposed for frequency-selective channels. In [5], C. Struder and G. Durisi formulated the *Maximum-A-Posteriori* (MAP) channel estimation and data detection for quantized systems as a convex optimization problem which can be efficiently solved using numerical methods. The achievable rates and equalization with a mix of low and high-resolution ADC architecture with perfect *Channel State Information* (CSI) is discussed in [6]. The computational complexity of the proposed methods [5], [6] are in general high. In [7], low-complexity linear channel estimation and receive combiners for multiuser symbol detection were proposed and a lower bound on capacity was also derived using linear schemes. To the best of our knowledge, all these results in the massive MIMO literature are limited only to CP-methods both for *Orthogonal Frequency Division Multiplexing* (OFDM) and *Single Carrier* (SC) transmission techniques, i.e., CP-OFDM and CP-SC.

The use of a CP enables efficient FDE using the *Fast Fourier Transform* (FFT). Moreover, it also offers flexibility on allocating time and frequency resources between sources for CP-OFDM. However, some of those advantages come at the price of a loss in spectral efficiency due to the CP. One of the goals of a massive MIMO system is to increase spectral efficiency. Therefore, it is pertinent to address the problem of equalization for the transmission schemes without the use of a CP.

Outside the realm of massive MIMO systems, the concept of FDE was proposed four decades ago to mitigate *InterSymbol Interference* (ISI) [8]. The idea has also been used by [9] in linear multiuser detection and downlink data detection in *Code Division Multiple Access* (CDMA) systems [10]. In [11], K. Hueske suggested an overlapping FDE that can be used as a pre-FFT equalizer in a CP-free OFDM transmission system. A *Sliding Window Frequency Domain Equalizer* (SW-FDE) was proposed for general multicarrier modulation receivers [12].

The idea of FDE applies a block-processing at the receiver side. It can be easily understood for a *Single Input Single Output* (SISO) system and has a straightforward extension for MIMO systems [9]. For a CP-free SISO system, the transmit symbols are linearly convolved with a channel impulse response and perturbed by white noise to arrive at the receiver-side. The receiver collects $N_b$ samples and does a model mismatch in the equalizer design by approximating a linear convolution with a circular convolution. The structure of the resulting equalizer can be efficiently implemented in the frequency domain using the FFT algorithm [13]. The equalization error distribution has a bathtub-like shape, which can be exploited by discarding the outer erroneous parts [11]. The number of symbols to be discarded and $N_b$ samples of a block length represent the trade-off between performance and computational complexity [14].

In this work we investigate the performance of FDE for massive MIMO systems deploying low resolution ADCs and operating over extremely frequency-selective channels. We formulate the optimal block length selection as a constraint *Optimization Problem* (OP), based on the computational complexity criteria. Interestingly, the optimal block length also achieves good performance in terms of BER- and MSE-criterion between transmitted and estimated symbols in addition to the lower computational complexity. Moreover, it is found that the optimal block length is strongly related to the channel impulse response and is independent of the ADC resolution and the number of receive antennas.

The paper is organized as follows. A quantized system model and a linearized version of it is introduced in Section II. Section IV describes the proposed efficient equalization scheme and discusses the computational complexity. In Section V we cast the optimal block length as a complexity optimization problem. Section VI solves the optimization problem and validates its performance using numerical simulations. Finally, Section VII concludes the paper.

Notation: Bold letters indicate vectors and matrices, non-bold letters express scalars. For a matrix $\boldsymbol{A}$, we denote complex conjugate, transpose and Hermitian transpose by $\boldsymbol{A}^*$, $\boldsymbol{A}^{\mathrm{T}}$ and $\boldsymbol{A}^{\mathrm{H}}$, respectively. The operator $\mathrm{diag}\left(\boldsymbol{A}\right)$ denotes a diagonal matrix containing only the diagonal elements of $\boldsymbol{A}$ and $\mathrm{vec}(\boldsymbol{A})$ denotes the vectorization operation transforming $\boldsymbol{A}$ into a column-vector by stacking all column vectors of $\boldsymbol{A}$ on top of each other. The expression $\boldsymbol{A} \otimes \boldsymbol{B}$ designates the Kronecker product between matrices $\boldsymbol{A}$ and $\boldsymbol{B}$. The $n \times n$ identity matrix is denoted by $\mathrm{I}_n$, while the zeros matrix with $n$ rows and $m$ columns is defined as $\boldsymbol{0}_{n\times m}$. We use $\mathrm{E}_{\boldsymbol{x}}\left[\cdot\right]$ to denote expectation with respect to the random vector $\boldsymbol{x}$.

## II. System Model

In the derivation of the system model an uplink of a single-cell scenario is considered, where the BS equipped with $M$ antennas receives the signals from $K$ single-antenna MSs. A frequency-selective block fading channel is assumed, which remains static for a coherence time of $T_c$ symbols between each pair of MS and BS antennas. In the following, we assume perfect CSI.

In the first subsection a quantized MIMO system model will be introduced for data detection. A linearized version of the quantized MIMO system model using Bussgang decomposition will be derived in the latter subsection.

### A. Quantized System Model

The channel between BS $m \in \{1, 2, \ldots, M\}$ and MS $k \in \{1, 2, \ldots, K\}$ is completely characterized by an impulse response of $L+1$ taps, denoted by $\boldsymbol{h}_{mk} \in \mathbb{C}^{L+1}$. The unquantized receive signal at BS $m$ is written as

$$y_m\left[n\right] = \sum_{l=0}^{L} \boldsymbol{h}_m^{\mathrm{T}}[l]\boldsymbol{x}\left[n-l\right] + \eta_m\left[n\right], \tag{1}$$

where $\boldsymbol{x}\left[n\right] = \begin{bmatrix} x_1\left[n\right] & x_2\left[n\right] & \cdots & x_K\left[n\right]\end{bmatrix}^{\mathrm{T}} \in \mathbb{C}^{K\times 1}$ is the zero-mean circularly-symmetric complex valued transmit vector with $\mathrm{E}_{\boldsymbol{x}}\left[\boldsymbol{x}\left[n\right]\cdot\boldsymbol{x}\left[n\right]^{\mathrm{H}}\right] = \sigma_x^2\boldsymbol{I}_K$ and $\boldsymbol{h}_m\left[l\right] = \begin{bmatrix} h_{m1}\left[l\right] & h_{m2}\left[l\right] & \cdots & h_{mK}\left[l\right]\end{bmatrix}^{\mathrm{T}} \in \mathbb{C}^{K\times 1}$ is constructed from the $l^{\mathrm{th}}$ tap of the channel impulse response from all users on the $m^{\mathrm{th}}$ antenna. Let the noise be drawn from the i.i.d. zero-mean circularly-symmetric complex Gaussian vector $\boldsymbol{\eta}\left[n\right] = \begin{bmatrix} \eta_1\left[n\right] & \eta_2\left[n\right] & \cdots & \eta_M\left[n\right]\end{bmatrix}^{\mathrm{T}} \in \mathbb{C}^{M\times 1}$, having the noise-covariance of $\mathrm{E}_{\boldsymbol{\eta}}\left[\boldsymbol{\eta}\left[n\right]\cdot\boldsymbol{\eta}\left[n\right]^{\mathrm{H}}\right] = \sigma_\eta^2\boldsymbol{I}_M$. We assume that the transmit and noise symbols are temporarily uncorrelated, i.e. $\mathrm{E}_{\boldsymbol{x}}\left[\boldsymbol{x}\left[n\right]\cdot\boldsymbol{x}\left[n-l\right]^{\mathrm{H}}\right] = 0$ for $l \neq 0$ and $\mathrm{E}_{\boldsymbol{\eta}}\left[\boldsymbol{\eta}\left[n\right]\cdot\boldsymbol{\eta}\left[n-l\right]^{\mathrm{H}}\right] = 0$ for $l \neq 0$. Using (1), the unquantized receive vector $\boldsymbol{y}\left[n\right] = \begin{bmatrix} y_1\left[n\right] & y_2\left[n\right] & \cdots & y_M\left[n\right]\end{bmatrix}^{\mathrm{T}} \in \mathbb{C}^{M\times 1}$ at time instant $n$ can be written as

$$\boldsymbol{y}[n] = \sum_{l=0}^{L}\boldsymbol{H}_l\boldsymbol{x}[n-l] + \boldsymbol{\eta}[n], \tag{2}$$

where $\boldsymbol{H}_l = \begin{bmatrix} \boldsymbol{h}_1\left[l\right] & \boldsymbol{h}_2\left[l\right] & \cdots & \boldsymbol{h}_M\left[l\right]\end{bmatrix}^{\mathrm{T}} \in \mathbb{C}^{M\times K}$ is a channel impulse response matrix. The signal vector $\boldsymbol{y}[n]$ is then quantized by a $b$-bit *uniform scalar* quantizer to obtain

$$\begin{aligned}\boldsymbol{r}[n] &= \begin{bmatrix}\mathrm{Q}_b\left(y_1\left[n\right]\right) & \mathrm{Q}_b\left(y_2\left[n\right]\right) & \cdots & \mathrm{Q}_b\left(y_M\left[n\right]\right)\end{bmatrix}^{\mathrm{T}}\\ &= \mathrm{Q}_b\left(\boldsymbol{y}[n]\right) = \mathrm{Q}_b\left(\sum_{l=0}^{L}\boldsymbol{H}_l\boldsymbol{x}[n-l] + \boldsymbol{\eta}[n]\right), \end{aligned}\tag{3}$$

where $\mathrm{Q}_b(.)$ represents the $b$-bit *uniform scalar* quantization operation and is applied element-wise to $\boldsymbol{y}[n]$, separately for the real and the imaginary part, i.e.,

$$r_{m,c}[n] = \mathrm{Q}_b\left(y_{m,c}[n]\right) = q^j_{m,c} \forall\ m \in \{1,\ldots,M\}, c \in \{\Re,\Im\}.$$

Here $q^j_{m,c}$ $\forall j \in \{1,\ldots,2^b\}$ represents the quantization level associated to the $j^{\mathrm{th}}$ quantization interval $]a_{j-1}, a_j]$. To cover a real valued input signal with arbitrary power, we set $a_0 = -\infty$ and $a_{2^b} = \infty$. The quantization levels are chosen to minimize the MSE of Gaussian signals as shown in [15, Table II].

Let us collect $N_b$ vectors, with a condition that $N_b > L$, corresponding to time instances $n, n-1, \ldots, n-(N_b-1)$ to form a space-time quantized receive matrix $\boldsymbol{R}[n]$, unquantized receive matrix $\boldsymbol{Y}[n]$, and noise matrix $\boldsymbol{N}[n]$ as

$$\begin{aligned}\boldsymbol{R}[n] &= \begin{bmatrix}\boldsymbol{r}\left[n\right] & \boldsymbol{r}\left[n-1\right] & \cdots & \boldsymbol{r}\left[n-(N_b-1)\right]\end{bmatrix} \in \mathbb{C}^{M\times N_b},\\ \boldsymbol{Y}[n] &= \begin{bmatrix}\boldsymbol{y}\left[n\right] & \boldsymbol{y}\left[n-1\right] & \cdots & \boldsymbol{y}\left[n-(N_b-1)\right]\end{bmatrix} \in \mathbb{C}^{M\times N_b},\\ \boldsymbol{N}[n] &= \begin{bmatrix}\boldsymbol{\eta}\left[n\right] & \boldsymbol{\eta}\left[n-1\right] & \cdots & \boldsymbol{\eta}\left[n-(N_b-1)\right]\end{bmatrix} \in \mathbb{C}^{M\times N_b}.\end{aligned}$$

As the channel impulse response has a memory of length $L$ (cf. Eq. (2)), the receive matrix $\boldsymbol{Y}[n]$ depends both on $\boldsymbol{X}_c[n] \in \mathbb{C}^{K\times N_b}$, i.e.,

$$\boldsymbol{X}_{\mathrm{c}}[n] = \begin{bmatrix}\boldsymbol{x}\left[n\right] & \boldsymbol{x}\left[n-1\right] & \cdots & \boldsymbol{x}\left[n-(N_b-1)\right]\end{bmatrix}, \tag{4}$$

and the interference matrix $\boldsymbol{X}_{\mathrm{in}}[n] \in \mathbb{C}^{K\times L}$ which consists of the previously transmitted vectors at time instances $n-N_b, \ldots, n-(N_b-1+L)$

$$\boldsymbol{X}_{\mathrm{in}}[n] = \begin{bmatrix}\boldsymbol{x}\left[n-N_b\right] & \cdots & \boldsymbol{x}\left[n-(N_b-1+L)\right]\end{bmatrix}. \tag{5}$$

The matrix $\boldsymbol{X}[n] \in \mathbb{C}^{K\times(N_b+L)}$ is then formed from $\boldsymbol{X}_{\mathrm{c}}[n]$ and $\boldsymbol{X}_{\mathrm{in}}[n]$

$$\boldsymbol{X}[n] = \begin{bmatrix}\boldsymbol{X}_{\mathrm{c}}\left[n\right] & \boldsymbol{X}_{\mathrm{in}}\left[n\right]\end{bmatrix}, \tag{6}$$

such that the space-time input-output relationship of the unquantized MIMO system is given as

$$\mathrm{vec}\{\boldsymbol{Y}[n]\} = \check{\boldsymbol{H}}\,\mathrm{vec}\{\boldsymbol{X}[n]\} + \mathrm{vec}\{\boldsymbol{N}[n]\}, \tag{7}$$

$$\check{\boldsymbol{y}}[n] = \check{\boldsymbol{H}}\check{\boldsymbol{x}}[n] + \check{\boldsymbol{\eta}}[n] \in \mathbb{C}^{M\cdot N_b\times 1} \text{ and} \tag{8}$$

$$\mathrm{vec}\{\boldsymbol{R}[n]\} = \check{\boldsymbol{r}}[n] = \mathrm{Q}_b\left(\check{\boldsymbol{y}}[n]\right) \in \mathbb{C}^{M\cdot N_b\times 1}, \tag{9}$$

where the channel matrix $\check{\boldsymbol{H}} \in \mathbb{C}^{M \cdot N_b \times K(N_b+L)}$ has a block Toeplitz structure of the form

$$\check{\boldsymbol{H}} = \begin{bmatrix} \boldsymbol{H}_0 & \boldsymbol{H}_1 & \cdots & \boldsymbol{H}_L & \boldsymbol{0} & \cdots & & \dots & \boldsymbol{0} \\ \boldsymbol{0} & \ddots & & & \ddots & \ddots & & & \vdots \\ & \ddots & \boldsymbol{H}_0 & \dots & \dots & \boldsymbol{H}_L & \boldsymbol{0} & & \\ & & & \boldsymbol{H}_0 & \dots & \boldsymbol{H}_L & \boldsymbol{H}_L & & \\ \vdots & & & & \ddots & \vdots & & \ddots & \\ \boldsymbol{0} & \dots & & & \boldsymbol{0} & \boldsymbol{H}_0 & \boldsymbol{H}_1 & \dots & \boldsymbol{H}_L \end{bmatrix}. \quad (10)$$

Here, the matrix $\boldsymbol{0}$ denotes $\boldsymbol{0}_{M \times K}$ for the sake of brevity.

## III. Linearized System Model Using Bussgang Decomposition

According to the Bussgang theorem [16], a nonlinear function such as a 1-bit quantizer with Gaussian input can be modeled as a linear transformation of the input signal $\boldsymbol{y}[n]$ and an additive distortion $\boldsymbol{e}[n]$ that is uncorrelated with the input, i.e.,

$$\check{\boldsymbol{r}}[n] = \mathrm{Q}_b\left(\check{\boldsymbol{y}}[n]\right) = \boldsymbol{B}\check{\boldsymbol{y}}[n] + \check{\boldsymbol{e}}[n]. \quad (11)$$

The expression for the matrix $\boldsymbol{B}$ and the covariance of the distortion $\check{\boldsymbol{e}}[n]$ are derived in [17] and are given as

$$\boldsymbol{B} = (1-\rho_q)\,\boldsymbol{I}_{M \cdot N_b} \text{ and } \boldsymbol{R}_{\check{\boldsymbol{e}}\check{\boldsymbol{e}}} = \rho_q\,(1-\rho_q)\,\mathrm{diag}\left(\boldsymbol{R}_{\check{\boldsymbol{y}}\check{\boldsymbol{y}}}\right), \quad (12)$$

where $\rho_q$ is the quantization distortion factor. The values of $\rho_q$ are listed in [15, Table II] and $\rho_q$ can be approximated by $\rho_q \approx 3^{-b}$. Using (11), the quantized MIMO system of (9) can be represented as an unquantized one:

$$\check{\boldsymbol{r}}[n] = (1-\rho_q)\,\check{\boldsymbol{H}}\check{\boldsymbol{x}}[n] + \check{\boldsymbol{\eta}}'[n] = \check{\boldsymbol{H}}'\check{\boldsymbol{x}}[n] + \check{\boldsymbol{\eta}}'[n], \quad (13)$$

where $\check{\boldsymbol{H}}' = (1-\rho_q)\,\check{\boldsymbol{H}}$ and

$$\begin{aligned} \boldsymbol{R}_{\check{\boldsymbol{\eta}}'\check{\boldsymbol{\eta}}'} &= (1-\rho_q)^2\,\boldsymbol{R}_{\check{\boldsymbol{\eta}}\check{\boldsymbol{\eta}}} + \rho_q\,(1-\rho_q)\,\mathrm{diag}\left(\boldsymbol{R}_{\check{\boldsymbol{y}}\check{\boldsymbol{y}}}\right) \\ &= (1-\rho_q)\left(\sigma_\eta^2 \boldsymbol{I}_{M \cdot N_b} + \rho_q\,\mathrm{diag}\left(\boldsymbol{I}_{N_b} \otimes \sum_{l=0}^{L} \boldsymbol{H}_l \boldsymbol{H}_l^{\mathrm{H}}\right)\right). \end{aligned} \quad (14)$$

## IV. Equalization Algorithm

In this section, an efficient FDE algorithm is introduced to estimate the transmitted data from the quantized observations $\check{\boldsymbol{r}}[n]$.

### A. Block-circulant Channel Approximation

The first step approximates the *block Toeplitz* channel matrix in the system model (13) as a *block-circulant* channel matrix. To this end, the receive signal $\check{\boldsymbol{r}}[n]$ can be expressed as a *superposition* of the product of the transmit vector $\check{\boldsymbol{x}}_{\mathrm{c}}[n] = \mathrm{vec}\{\boldsymbol{X}_{\mathrm{c}}[n]\}$ (cf. Eq. (4)) with a channel matrix $\check{\boldsymbol{H}}'_{\mathrm{c}}$

$$\check{\boldsymbol{H}}'_{\mathrm{c}} = (1-\rho_q) \begin{bmatrix} \boldsymbol{H}_0 & \boldsymbol{H}_1 & \cdots & \boldsymbol{H}_L & \boldsymbol{0} & \cdots \\ & \ddots & & & \ddots & \\ & & \boldsymbol{H}_0 & \dots & \dots & \boldsymbol{H}_L \\ & & & \boldsymbol{H}_0 & \dots & \boldsymbol{H}_L \\ \vdots & \ddots & & & \ddots & \vdots \\ \boldsymbol{0} & \dots & & & & \boldsymbol{H}_0 \end{bmatrix}, \text{ and} \quad (15)$$

an interfering vector $\check{\boldsymbol{z}}_{\mathrm{in}}[n] = \left[\mathrm{vec}\{\boldsymbol{X}_{\mathrm{in}}[n]\}^{\mathrm{T}} \;\; \boldsymbol{0}_{1 \times (N_b - L)K}\right]^{\mathrm{T}}$ (cf. Eq. (5)) with a channel matrix $\check{\boldsymbol{H}}'_{\mathrm{in}}$

$$\check{\boldsymbol{H}}'_{\mathrm{in}} = (1-\rho_q) \begin{bmatrix} \boldsymbol{0} & \boldsymbol{0} & \cdots & \boldsymbol{0} & \boldsymbol{0} & \cdots \\ & \ddots & & & \ddots & \\ & & \boldsymbol{0} & \dots & \dots & \boldsymbol{0} \\ \boldsymbol{H}_L & & \boldsymbol{0} & & \dots & \boldsymbol{0} \\ \vdots & \ddots & & & \ddots & \vdots \\ \boldsymbol{H}_1 & \dots & \boldsymbol{H}_L & & & \boldsymbol{0} \end{bmatrix}, \text{ such that} \quad (16)$$

$$\begin{aligned} \check{\boldsymbol{r}}[n] &= \check{\boldsymbol{H}}'_{\mathrm{c}}\check{\boldsymbol{x}}_{\mathrm{c}}[n] + \check{\boldsymbol{\eta}}'[n] + \check{\boldsymbol{H}}'_{\mathrm{in}}\check{\boldsymbol{z}}_{\mathrm{in}}[n] \\ &= \boldsymbol{H}_{\mathrm{cir}}\check{\boldsymbol{x}}_{\mathrm{c}}[n] + \check{\boldsymbol{\eta}}'[n] + \check{\boldsymbol{\gamma}}'[n]. \end{aligned} \quad (17)$$

In (17), $\check{\boldsymbol{\gamma}}'[n] = \check{\boldsymbol{H}}'_{\mathrm{in}}\left(\check{\boldsymbol{z}}_{\mathrm{in}}[n] - \check{\boldsymbol{x}}_{\mathrm{c}}[n]\right)$ can be considered as an interference noise which corrupts the last $M \cdot L$ samples of $\check{\boldsymbol{r}}[n]$ and $\boldsymbol{H}_{\mathrm{cir}} \in \mathbb{C}^{M \cdot N_b \times K \cdot N_b}$

$$\boldsymbol{H}_{\mathrm{cir}} = \check{\boldsymbol{H}}'_{\mathrm{c}} + \check{\boldsymbol{H}}'_{\mathrm{in}} = \begin{bmatrix} \boldsymbol{H}_0 & \boldsymbol{H}_1 & \cdots & \boldsymbol{H}_L & \boldsymbol{0} & \cdots \\ & \ddots & & & \ddots & \\ & & \boldsymbol{H}_0 & \dots & \dots & \boldsymbol{H}_L \\ \boldsymbol{H}_L & & & \boldsymbol{H}_0 & \dots & \boldsymbol{H}_L \\ \vdots & \ddots & & & \ddots & \vdots \\ \boldsymbol{H}_1 & \dots & \boldsymbol{H}_L & & & \boldsymbol{H}_0 \end{bmatrix}, \quad (18)$$

is a *block-circulant* matrix. The linear *Minimum Mean Square Error* (MMSE)-estimator or *Wiener Filter* (WF) $\check{\boldsymbol{G}}$ is applied to $\check{\boldsymbol{r}}[n]$ in (17) to get $\hat{\check{\boldsymbol{x}}}_c[n]$, i.e.,

$$\hat{\check{\boldsymbol{x}}}_c[n] = \check{\boldsymbol{G}}\check{\boldsymbol{r}}[n] = \left(\boldsymbol{H}_{\mathrm{cir}}^{\mathrm{H}}\boldsymbol{R}_{\check{\boldsymbol{\eta}}'\check{\boldsymbol{\eta}}'}^{-1}\boldsymbol{H}_{\mathrm{cir}} + \boldsymbol{R}_{\check{\boldsymbol{x}}'\check{\boldsymbol{x}}'}^{-1}\right)^{-1}\boldsymbol{H}_{\mathrm{cir}}^{\mathrm{H}}\boldsymbol{R}_{\check{\boldsymbol{\eta}}'\check{\boldsymbol{\eta}}'}^{-1}\check{\boldsymbol{r}}[n]. \quad (19)$$

It is important to mention that the term $\check{\boldsymbol{G}}\check{\boldsymbol{\gamma}}'[n]$ is an *interference distortion* which corrupts the whole estimated data-block $\hat{\check{\boldsymbol{x}}}_c[n]$. The next subsection presents a methodology to minimize the interference distortion.

### B. Interference Analyses

It was shown in [11] that the ensemble-averaged *interference distortion* power has a bathtub like distribution. This behavior can be exploited to minimize the resulting error by using a $L'$ samples overlapping of data blocks, i.e., $\boldsymbol{R}[n]$ contains vectors corresponding to the time instances $n, \dots, n-(N_b-1)$ and is followed by $\boldsymbol{R}[n-(N_b-L')]$ with corresponding time elements $n-(N_b-L'), \dots, n-(2N_b-L'-1)$.

The value of the overlap $L'$ is a design parameter and experiments in the literature take it for instance as 50% [12] or 30% or 60% [18] of the block length $N_b$. In this work the value of $L'$ is fixed and is directly related to the length of the channel memory [14], i.e., $L' = L$. Since $L$ can be even or odd, we define the pre-discard- and post-discard-lengths as $L_{\mathrm{pre}} = \lceil L'/2 \rceil$ and $L_{\mathrm{post}} = \lfloor L'/2 \rfloor$, respectively, such that $L_{\mathrm{pre}} + L_{\mathrm{post}} = L' = L$.

### C. Frequency Domain Equalization (FDE)

We will now give the methodology to solve (19) efficiently in the frequency domain. A block-circulant matrix can be block-diagonalized as

$$\boldsymbol{\mathcal{H}} = \left(\boldsymbol{F} \otimes \boldsymbol{I}_M\right)\boldsymbol{H}_{\mathrm{cir}}\left(\boldsymbol{F}^{\mathrm{H}} \otimes \boldsymbol{I}_K\right). \quad (20)$$

The matrix $\boldsymbol{\mathcal{H}}$ is now block-diagonal:

$$\boldsymbol{\mathcal{H}} = \mathrm{diag}\{\boldsymbol{H}_{f_i}\}_{i=1}^{N_b}, \text{ where} \tag{21}$$

$$\boldsymbol{H}_{f_i} = (1-\rho_q)\sum_{l=0}^{L}\boldsymbol{H}_l \cdot \exp\left(-\mathrm{j}\cdot\frac{2\pi}{N_b}l(i-1)\right), \text{for } 1\leq i\leq N_b,$$

are channel matrices in the *frequency domain representation* of the multipath MIMO channel. Ignoring the interference term $\check{\boldsymbol{\gamma}}'[n]$ in (17) and multiplying it from the left with $(\boldsymbol{F}\otimes\boldsymbol{I}_M)$, we obtain:

$$\begin{aligned}(\boldsymbol{F}\otimes\boldsymbol{I}_M)\check{\boldsymbol{r}}[n] &\approx (\boldsymbol{F}\otimes\boldsymbol{I}_M)\,\boldsymbol{H}_{\mathrm{cir}}\check{\boldsymbol{x}}_{\mathrm{c}}[n] + (\boldsymbol{F}\otimes\boldsymbol{I}_M)\,\check{\boldsymbol{\eta}}'[n] \\ &\overset{(20)}{=} \boldsymbol{\mathcal{H}}\,(\boldsymbol{F}\otimes\boldsymbol{I}_K)\,\check{\boldsymbol{x}}_{\mathrm{c}}[n] + (\boldsymbol{F}\otimes\boldsymbol{I}_M)\,\check{\boldsymbol{\eta}}'[n].\end{aligned} \tag{22}$$

The noise statistics in frequency domain remain unchanged, such that $(\boldsymbol{F}\otimes\boldsymbol{I}_M)\,\boldsymbol{R}_{\check{\boldsymbol{\eta}}'\check{\boldsymbol{\eta}}'}\left(\boldsymbol{F}^{\mathrm{H}}\otimes\boldsymbol{I}_M\right) = \boldsymbol{R}_{\check{\boldsymbol{\eta}}'\check{\boldsymbol{\eta}}'}$ Using the vectorization operator, we get

$$\begin{aligned}(\boldsymbol{F}\otimes\boldsymbol{I}_M)\,\mathrm{vec}\{\boldsymbol{R}[n]\} &= \boldsymbol{\mathcal{H}}\,(\boldsymbol{F}\otimes\boldsymbol{I}_K)\,\mathrm{vec}\{\boldsymbol{X}_{\boldsymbol{c}}[n]\} \\ &\quad + (\boldsymbol{F}\otimes\boldsymbol{I}_M)\,\mathrm{vec}\{\boldsymbol{N'}[n]\}, \\ \mathrm{vec}\{\boldsymbol{R}[n]\boldsymbol{F}^{\mathrm{T}}\} &= \boldsymbol{\mathcal{H}}\,\mathrm{vec}\{\boldsymbol{X}_{\boldsymbol{c}}[n]\boldsymbol{F}^{\mathrm{T}}\} + \mathrm{vec}\{\boldsymbol{N'}[n]\boldsymbol{F}^{\mathrm{T}}\},\end{aligned} \tag{23}$$

where the tensor equality $(\boldsymbol{C}\otimes\boldsymbol{A})\,\mathrm{vec}\{\boldsymbol{D}\} = \mathrm{vec}\{\boldsymbol{ADC}^{\mathrm{T}}\}$ was applied and $\boldsymbol{N'}[n]$ was formed by reshaping $\check{\boldsymbol{\eta}}'$ (cf. $\boldsymbol{N}[n]$ in Sec. II). Let us define the frequency domain matrices:

$$\boldsymbol{R}_{\boldsymbol{f}} = \boldsymbol{R}[n]\boldsymbol{F}^{\mathrm{T}} = \begin{bmatrix}\boldsymbol{r}_{f_0} & \boldsymbol{r}_{f_1} & \cdots & \boldsymbol{r}_{f_{N_b}}\end{bmatrix} \in \mathbb{C}^{M\times N_b}, \tag{24}$$

$$\boldsymbol{X}_{\boldsymbol{f}} = \boldsymbol{X}[n]\boldsymbol{F}^{\mathrm{T}} = \begin{bmatrix}\boldsymbol{x}_{f_1} & \boldsymbol{x}_{f_2} & \cdots & \boldsymbol{x}_{f_{N_b}}\end{bmatrix} \in \mathbb{C}^{K\times N_b}, \tag{25}$$

$$\boldsymbol{N}_{\boldsymbol{f}} = \boldsymbol{N'}[n]\boldsymbol{F}^{\mathrm{T}} = \begin{bmatrix}\boldsymbol{\eta}_{f_1} & \boldsymbol{\eta}_{f_2} & \cdots & \boldsymbol{\eta}_{f_{N_b}}\end{bmatrix} \in \mathbb{C}^{M\times N_b}. \tag{26}$$

The rows of $\boldsymbol{R}_{\boldsymbol{f}}$, $\boldsymbol{X}_{\boldsymbol{f}}$, $\boldsymbol{N}_{\boldsymbol{f}}$ are the Fourier transform of the rows of $\boldsymbol{R}[n]$, $\boldsymbol{X}_{\mathrm{c}}[n]$ and $\boldsymbol{N'}[n]$, respectively, and the $i^{\mathrm{th}}$ column of the matrices represents the frequency domain description of the $i^{\mathrm{th}}$ frequency subband. The input-output relationship of the $i^{\mathrm{th}}$ frequency band is given as

$$\boldsymbol{r}_{f_i} = \boldsymbol{H}_{f_i}\boldsymbol{x}_{f_i} + \boldsymbol{\eta}_{f_i}, \text{for } 1\leq i\leq N_b. \tag{27}$$

The MMSE equalizer is applied in each subband to estimate

$$\begin{aligned}\widehat{\boldsymbol{x}}_{f_i} &= \boldsymbol{G}_{f_i}\boldsymbol{r}_{f_i} \\ &= \left(\boldsymbol{H}_{f_i}^{\mathrm{H}}\boldsymbol{R}_{\boldsymbol{\eta}_{f_i}\boldsymbol{\eta}_{f_i}}^{-1}\boldsymbol{H}_{f_i} + \boldsymbol{R}_{\boldsymbol{x}_{f_i}\boldsymbol{x}_{f_i}}^{-1}\right)^{-1}\boldsymbol{H}_{f_i}^{\mathrm{H}}\boldsymbol{R}_{\boldsymbol{\eta}_{f_i}\boldsymbol{\eta}_{f_i}}^{-1}\boldsymbol{r}_{f_i}.\end{aligned} \tag{28}$$

Using (28), the estimated frequency matrix $\widehat{\boldsymbol{X}}_f = \begin{bmatrix}\widehat{\boldsymbol{x}}_{f_1} & \widehat{\boldsymbol{x}}_{f_2} & \cdots & \widehat{\boldsymbol{x}}_{f_{N_b}}\end{bmatrix} \in \mathbb{C}^{K\times N_b}$ is converted back into time by the inverse Fourier transform, i.e.,

$$\widehat{\boldsymbol{X}}_c[n] = \widehat{\boldsymbol{X}}_f\boldsymbol{F}^* = \begin{bmatrix}\widehat{\boldsymbol{x}}[n] & \widehat{\boldsymbol{x}}[n-1] & \cdots & \widehat{\boldsymbol{x}}[n-(N_b-1)]\end{bmatrix}. \tag{29}$$

### *D. Computational Complexity*

The computational complexity of the FDE involves taking the $N_b$-point FFT of the receive symbols on each antenna (24), calculating the frequency domain WF for each frequency subband (28) and an $N_b$-point *Inverse Fast Fourier Transform* (IFFT) to convert the estimated symbols for each user (29) back into time domain. The total number of complex multiplications to process $N_b$ symbols for all $K$ users are

$$\mathcal{T}_{\mathrm{d}}(N_b) = (M+K)\cdot N_b\cdot\log_2 N_b + K\cdot M\cdot N_b. \tag{30}$$

It is assumed that the WFs $\boldsymbol{G}_{f_i}$ for all the subbands are pre-computed once for the whole coherence time $T_c$. This cost is not taken into consideration in (30) and will be stated in the next subsection along with an optimization problem for obtaining the optimal block length.

## V. Computational Complexity Optimization

The computational complexity of the FDE with overlap-discard processing involves a static-part $\mathcal{T}_{\mathrm{s}}(N_b)$, which is to be computed once per coherence time and a dynamic-part $\mathcal{T}_{\mathrm{d}}(N_b)$ (30), which has to be calculated for each frame of length $N_b$. In the following analysis we assume the block length to be a power of 2 such that we can employ the efficient FFT-algorithm.

The static cost $\mathcal{T}_{\mathrm{s}}(N_b)$ incorporates the $N_b$-point FFT of the $KM$ channel impulse response vectors (21) and calculating the WF $\boldsymbol{G}_{f_i}$ for each subband (28).

Since $(\boldsymbol{F}\otimes\boldsymbol{I}_M)\,\boldsymbol{R}_{\check{\boldsymbol{\eta}}'\check{\boldsymbol{\eta}}'}\left(\boldsymbol{F}^{\mathrm{H}}\otimes\boldsymbol{I}_M\right) = \boldsymbol{R}_{\check{\boldsymbol{\eta}}'\check{\boldsymbol{\eta}}'}$ is a diagonal matrix (cf. Eq. (14)), $\boldsymbol{R}_{\boldsymbol{\eta}_{f_i}\boldsymbol{\eta}_{f_i}}$ is also a diagonal matrix. Moreover, $\boldsymbol{R}_{\boldsymbol{x}_{f_i}\boldsymbol{x}_{f_i}}$ is a scaled identity-matrix, i.e. $\sigma_x^2\mathrm{I}_K$. Since inversion of $\boldsymbol{R}_{\boldsymbol{\eta}_{f_i}\boldsymbol{\eta}_{f_i}}$ is inexpensive and adding an inverse identity-matrix $\boldsymbol{R}_{\boldsymbol{x}_{f_i}\boldsymbol{x}_{f_i}}^{-1}$ in (28) only involves additions both steps are neglected. Accordingly, $\mathcal{T}_{\mathrm{s}}(N_b)$ is given as

$$\mathcal{T}_{\mathrm{s}}(N_b) = \underbrace{\left(\underbrace{KM\log_2(N_b)}_{(a)} + \underbrace{KM}_{(b)} + \underbrace{2\cdot K^2M}_{(c+e)} + \underbrace{K^3}_{(d)}\right)\cdot N_b}_{(f)}, \tag{31}$$

which consists of converting the $KM$ channel impulse responses bin-wise to the frequency domain (21) $(a)$, performing the matrix-matrix multiplication $\boldsymbol{H}_{f_i}^{\mathrm{H}}\boldsymbol{R}_{\boldsymbol{\eta}_{f_i}\boldsymbol{\eta}_{f_i}}^{-1}$ $(b)$ and saving the result as $\boldsymbol{O}$, multiplying $\boldsymbol{OH}_{f_i}$ $(c)$, doing an inversion on the resulting $K\times K$ matrix $(d)$ and saving it as $\boldsymbol{P}$, evaluating $\boldsymbol{PO}$ $(e)$ and finally multiplying the obtained complexity with $N_b$ to account for all subbands $(f)$.

In accordance to that, we derive the total computational complexity $\mathcal{T}_{\mathrm{tot}}(N_b)$, taking the entire coherence time $T_c$, as

$$\mathcal{T}_{\mathrm{tot}}(N_b) = \mathcal{T}_{\mathrm{s}}(N_b) + \mathcal{T}_{\mathrm{d}}(N_b)\cdot\frac{T_c}{N_b - L'}, \tag{32}$$

where $T_c/(N_b - L')$ describes the number of frames processed per coherence time using the underlying overlap-and-discard scheme. In order to arrive at the total cost per estimated symbol $\mathcal{T}_{\mathrm{sym}}(N_b)$, we will divide (32) by $KT_c$, i.e. the total amount of transmitted symbols per coherence time of all the $K$ users and arrive at:

$$\begin{aligned}\mathcal{T}_{\mathrm{sym}}(N_b) &= \frac{\mathcal{T}_{\mathrm{tot}}(N_b)}{KT_c} \\ &= \frac{\mathcal{T}_{\mathrm{s}}(N_b)}{KT_c} + \frac{\mathcal{T}_{\mathrm{d}}(N_b)}{K(N_b - L')}.\end{aligned} \tag{33}$$

In order to obtain the minimal cost per estimated symbol, we are interested in the optimizer $N_b^{(\mathrm{opt})}$, which minimizes (33) given a certain constraint set. The constraint optimization problem is formulated as follows:

$$\begin{aligned} N_b^{(\mathrm{opt})} = \underset{N_b}{\operatorname{argmin}} \quad & \mathcal{T}_{\mathrm{sym}}(N_b) \\ \text{subject to} \quad & N_b \le T_c, \\ & N_b \ge (L'+1). \end{aligned} \tag{34}$$

The first constraint ensures that the optimal block length is smaller than or equal to $T_c$, whereas the second constraint limits the solution space to block lengths greater than or equal to $L' + 1$, assuring a minimum retain of *one* sample per estimated block in the overlap-save processing.

The optimization problem in (34) does not have a closed-form solution and a numerical procedure has been applied in the next section to solve for $N_b^{(\mathrm{opt})}$. Furthermore, it will be shown that $N_b^{(\mathrm{opt})}$ also gives a reasonable good performance in terms of MSE and BER for a given MIMO setup.

## VI. Simulation Results and Analysis

Consider a MIMO setup having $K$ users with a single transmit antenna and $M$ receive antennas. The channel impulse response between each pair of transmit and receive antennas consists of $L + 1$ taps.

The transmitter is employing 16-QAM *(Quadrature Amplitude Modulation)*, CP is omitted at the transmit side and the receiver is using 1-bit quantizers at each of the receive antennas. It is assumed that the coherence time of the channel is $T_c = 50 \times 10^3$ symbols, if not otherwise denoted. The noise is zero-mean circularly-symmetric *Additive White Gaussian Noise* (AWGN) with variance $\sigma_\eta^2 = 1$ (cf. Sec. II) and the channel is chosen based on an Extended Vehicular A model (9 nonzero taps). The results are averaged over $N_{\mathrm{sim}} = 200$ channel realizations. Perfect CSI is assumed throughout the data estimation process. The bit energy to noise spectral density $E_b/N_0$ is defined as:

$$\frac{E_b}{N_0} = \frac{P_t}{KM\sigma_\eta^2} \frac{\operatorname{trace}\left(\mathrm{E}_{\check{\boldsymbol{H}}}\left(\check{\boldsymbol{H}}\check{\boldsymbol{H}}^H\right)\right)}{B}, \tag{35}$$

where $P_t$ is the total transmit power, $B$ is the number of bits per constellation symbol and $\check{\boldsymbol{H}}$ is the linear channel convolution matrix (cf. Eq. (10)). In the data estimation the MSE is defined as

$$\mathrm{MSE} = \frac{1}{N_{\mathrm{sim}} K T_c} \sum_{n_{\mathrm{sim}}=1}^{N_{\mathrm{sim}}} \sum_{k=1}^{K} \left\| \hat{\boldsymbol{x}}_k^{(n_{\mathrm{sim}})} - \boldsymbol{x}_k^{(n_{\mathrm{sim}})} \right\|_2^2, \tag{36}$$

where $\boldsymbol{x}_k^{(n_{\mathrm{sim}})}$ is the entire transmit vector of the $k^{\mathrm{th}}$ user, consisting of $T_c$ symbols per realization $n_{\mathrm{sim}}$, and $\hat{\boldsymbol{x}}_k^{(n_{\mathrm{sim}})}$ is its according estimate. The receiver does not employ blind constellation power scaling.

FDE with overlap-save processing is assumed throughout this section and we discard $L' = L$ samples from each estimated block of length $N_b$ (cf. Sec. IV-B). The MSE- and BER-criterion is taken as a performance measure for data-detection based on different linear estimation methods and block lengths for equalization.

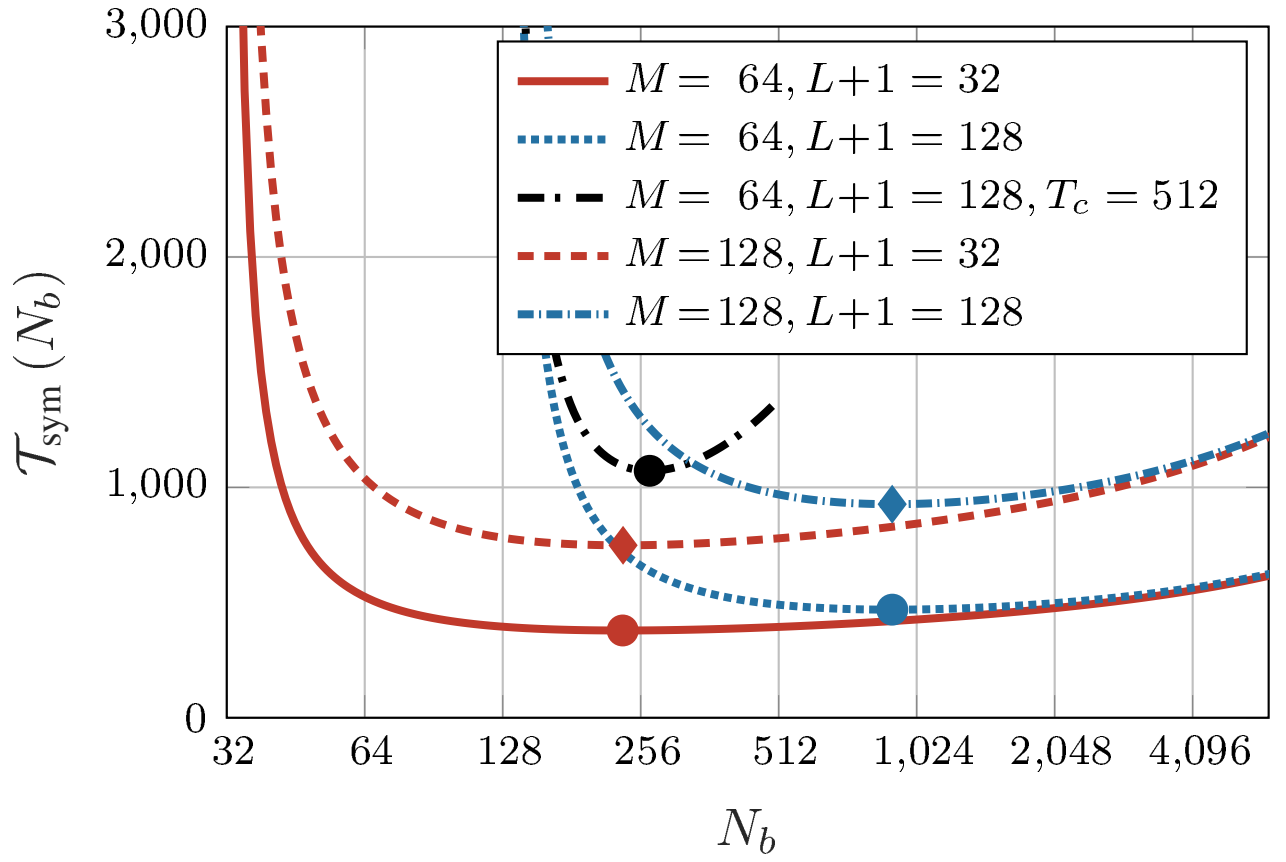

Fig. 1. Computational complexity per estimated symbol as a function of block length $N_b$ for $K = 2$ users, $M = 64, 128$ receive antennas and channel impulse response lengths $L + 1 = 32, 128$. The minimal complexity $N_b^{(\mathrm{opt})}$ is depicted with a dot-mark ($M = 64$) and a diamond-mark ($M = 128$). $T_c$ is assumed to be $50 \times 10^3$ symbols if not otherwise denoted.

*First Experiment - The optimal block length:* The first experiment in Fig. 1 shows the computational complexity per estimated symbol $\mathcal{T}_{\mathrm{sym}}(N_b)$ based on (33) as a function of block length $N_b$ for $M = 64, 128$ receive antennas and $L + 1 = 32, 128$ channel impulse response taps. The first important observation is that choosing $N_b$ close to $L' + 1$ becomes very costly, as only very few samples will be retained per processed block, resulting in a huge number of frames per $T_c$ to be processed. On the other hand, choosing a large $N_b$ will also come with very high complexity, as $\mathcal{T}_{\mathrm{s}}(N_b)$ and $\mathcal{T}_{\mathrm{d}}(N_b)$ increase with $N_b \log_2(N_b)$, although now only fewer frames per $T_c$ will need to be processed. Obviously, for a smaller coherence time, the optimal block length will also become smaller, as the first term in (33) will now contribute significantly more to the complexity per symbol. Another important observation comprises that scaling the number of receive antennas only offsets the complexity, but leaves the optimal value $N_b^{(\mathrm{opt})}$ unchanged. Furthermore, increasing the channel impulse response length $L$, the optimal block length for FDE becomes larger.

Therefore, the optimization of $\mathcal{T}_{\mathrm{sym}}(N_b)$ describes a trade-off between processing large blocks (*costly*, cf. Eqns. (30), (31)) and handling few of those blocks per $T_c$ (*cheap*), and processing small blocks (*cheap*, cf. Eqns. (30), (31)) and handling more of those blocks per $T_c$ (*costly*).

From an implementation-wise point of view, we need to take block lengths being a power of 2 in order for the assumptions in (33) to hold. However, this poses no problem, as either the prior or the next power of 2 from $N_b^{(\mathrm{opt})}$ is insignificantly more costly.

*Second Experiment - Performance of the optimal block length:* The next experiments in Fig. 2 and 3 compare the performance of FDE with the overlap-save scheme in terms of MSE and BER as a function of block length and $E_b/N_0$. The setup comprises $K = 2$ users, $M = 64$ receive antennas and $L + 1 = 128$ channel impulse response taps. Two FDE approaches are used to estimate the transmit symbols.

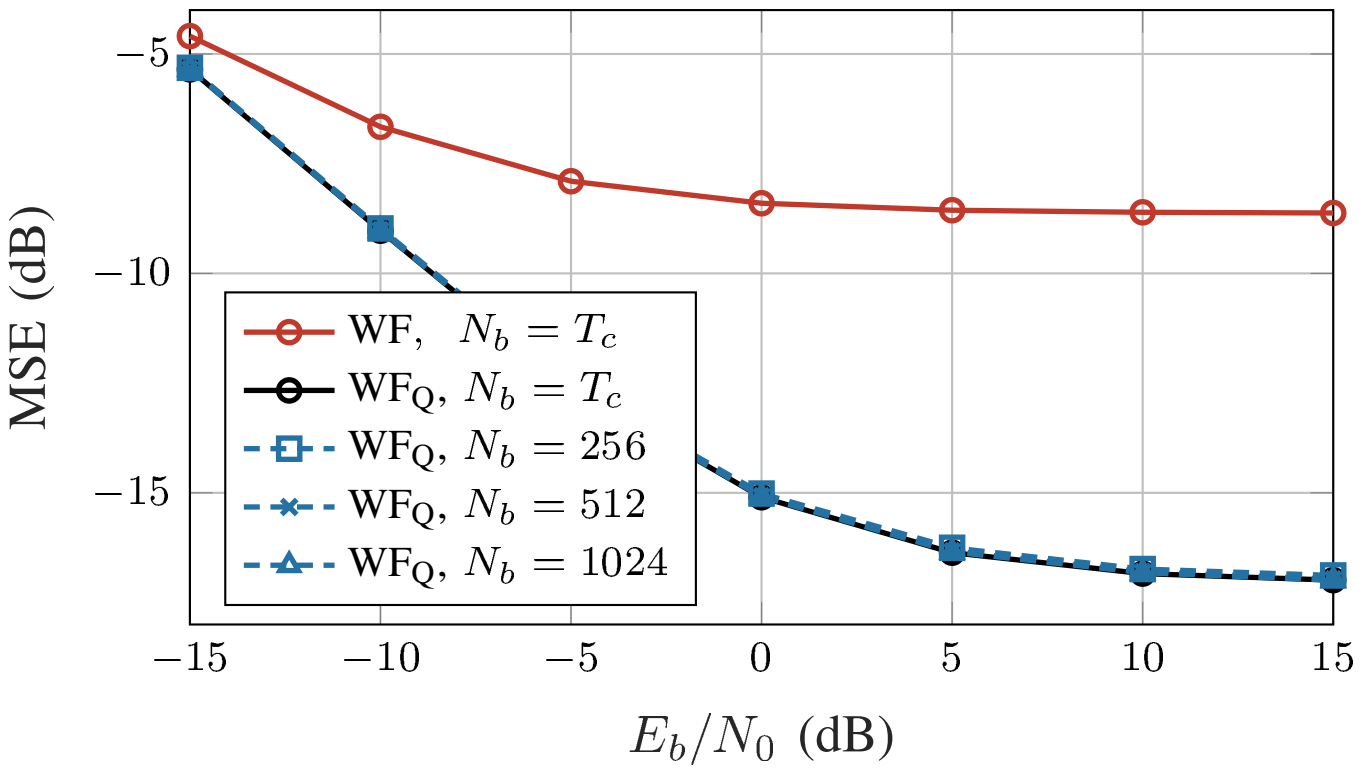

Fig. 2. MSE with neglected quantization perturbation (WF) and incorporated quantization perturbation (WF$_{\mathrm{Q}}$) as a function of $E_b/N_0$ and different block lengths $N_b$ (plotmarks). $K = 2$ users, $M = 64$ receive antennas, channel impulse response length $L + 1 = 128$.

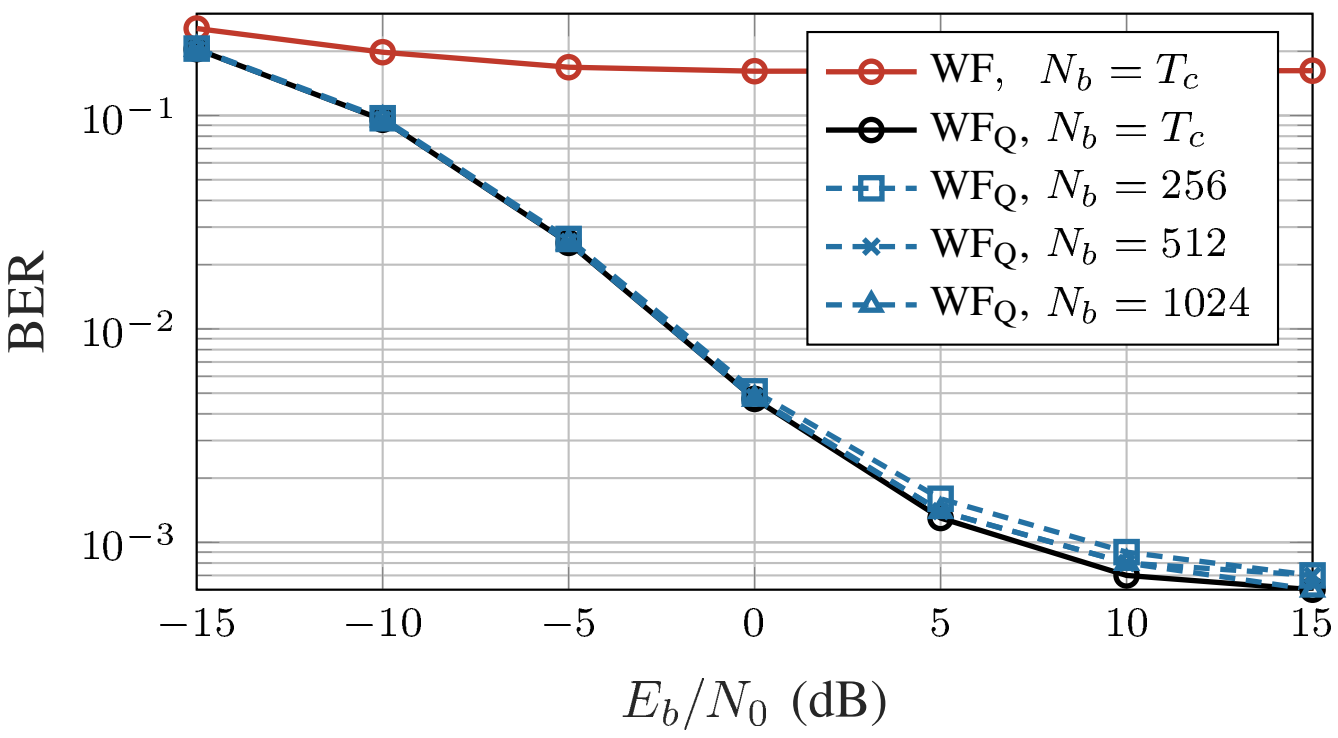

Fig. 3. Uncoded BER with neglected quantization perturbation (WF) and incorporated quantization perturbation (WF$_{\mathrm{Q}}$) as a function of $E_b/N_0$ and different block lengths $N_b$ (plotmarks). $K = 2$ users, $M = 64$ receive antennas, channel impulse response length $L + 1 = 128$.

WF is a Wiener Filter approach neglecting the quantization perturbation (cf. Eq. (13), $\rho_q = 0$), whereas WF$_{\mathrm{Q}}$ incorporates the quantization perturbation using the Bussgang Theorem (cf. Eq. (28), $\rho_q \neq 0$). Evidently, a performance increase in terms of the MSE and BER is observable when comparing the two estimation methods WF and WF$_{\mathrm{Q}}$, as the latter takes the quantization disturbance into account. Both metrics furthermore conclude that taking the optimal block length $N_b^{(\mathrm{opt})}$ closely matches the performance of equalizing all symbols per coherence time $T_c$ at the same time, i.e. having no inter-symbol-interference and taking the maximum block length $N_b = T_c$ (cf. Eq. (34)). This observation holds also for different $E_b/N_0$ levels. Deviating from this optimal block length will in either case increase computational cost, but only increase data-detection performance to an insignificant extend.

## VII. Conclusion

In this paper we have presented an efficient block-processing data-estimation approach for CP-free, quantized massive MIMO systems having frequency-selective channels between the MSs and the BS. A complexity optimization problem for obtaining the optimal block length has been stated. Although the optimal block is dervied based on the linearized approximation of quantization, the simulation results verfiy that the optimal block length minimizes the MSE and BER performance criterion. Additionally, the reduced computational complexity by the optimal block length doesn't degrade the performance compared with processing all the sysmbols during the coherence time, i.e., $N_b = T_c$ although the optimal block length has a reduced complexity with the latter one.

There are many avenues for the future work. In our analysis we assume perfect CSI, and extending the proposed approach to the imperfect CSI is an interesting direction. Future work of interest could also focus on including the value of overlapping $L'$ into the complexity optimization cost function.